\documentclass[nonacm,sigconf]{acmart}

\newcommand{\ourtool}{StreamFunnel }
\newcommand{\ourtooln}{StreamFunnel}

\renewcommand\footnotetextcopyrightpermission[1]{}
\setcopyright{none}
\begin{document}

\title{\ourtooln}
\subtitle{Facilitating communication between a VR streamer and many spectators}

\settopmatter{authorsperrow=4}
\author{Haohua Lyu}
\authornote{The authors contributed equally to this research.}
\affiliation{%
    \institution{University of California, Berkeley}
    \city{Berkeley}
    \state{CA}
    \country{USA}
}
\email{haohua@berkeley.edu}

\author{Cyrus Vachha}
\authornotemark[1]
\affiliation{%
    \institution{University of California, Berkeley}
    \city{Berkeley}
    \state{CA}
    \country{USA}
}
\email{cvachha@berkeley.edu}

\author{Qianyi Chen}
\authornotemark[1]
\affiliation{%
    \institution{University of California, Berkeley}
    \city{Berkeley}
    \state{CA}
    \country{USA}
}
\email{qianyi.chen@berkeley.edu}

\author{Balasaravanan Thoravi Kumaravel}
\affiliation{%
    \institution{University of California, Berkeley}
    \city{Berkeley}
    \state{CA}
    \country{USA}
}
\email{bala@eecs.berkeley.edu}

\author{Bj\"{o}ern Hartmann}
\affiliation{%
    \institution{University of California, Berkeley}
    \city{Berkeley}
    \state{CA}
    \country{USA}
}
\email{bjoern@eecs.berkeley.edu}

\renewcommand{\shortauthors}{Haohua Lyu, Cyrus Vachha, Qianyi Chen, et al.}

\renewcommand{\shortauthors}{\ourtooln}

\begin{abstract}

The increasing adoption of Virtual Reality (VR) systems in different domains have led to a need to support interaction between many spectators and a VR user. This is common in game streaming, live performances, and webinars. Prior CSCW systems for VR environments are limited to small groups of users. In this work, we identify problems associated with interaction carried out with large groups of users. To address this, we introduce an additional user role — the co-host. They mediate communications between the VR user and  many spectators. To facilitate this mediation, we present \ourtooln, which allows the co-host to be part of the VR application's space and interact with it. The design of \ourtool was informed by formative interviews with six experts. \ourtool uses a cloud-based streaming solution to enable remote co-host and many spectators to view and interact through standard web browsers, without requiring any custom software. We present results of informal user testing which provides insights into \ourtooln's ability to facilitate these scalable interactions. Our participants, who took the role of a co-host, found that \ourtool enables them to add value in presenting the VR experience to the spectators and relaying useful information from the live chat to the VR user.  

\end{abstract}



\begin{teaserfigure}
  \centering
  \includegraphics[scale=0.18]{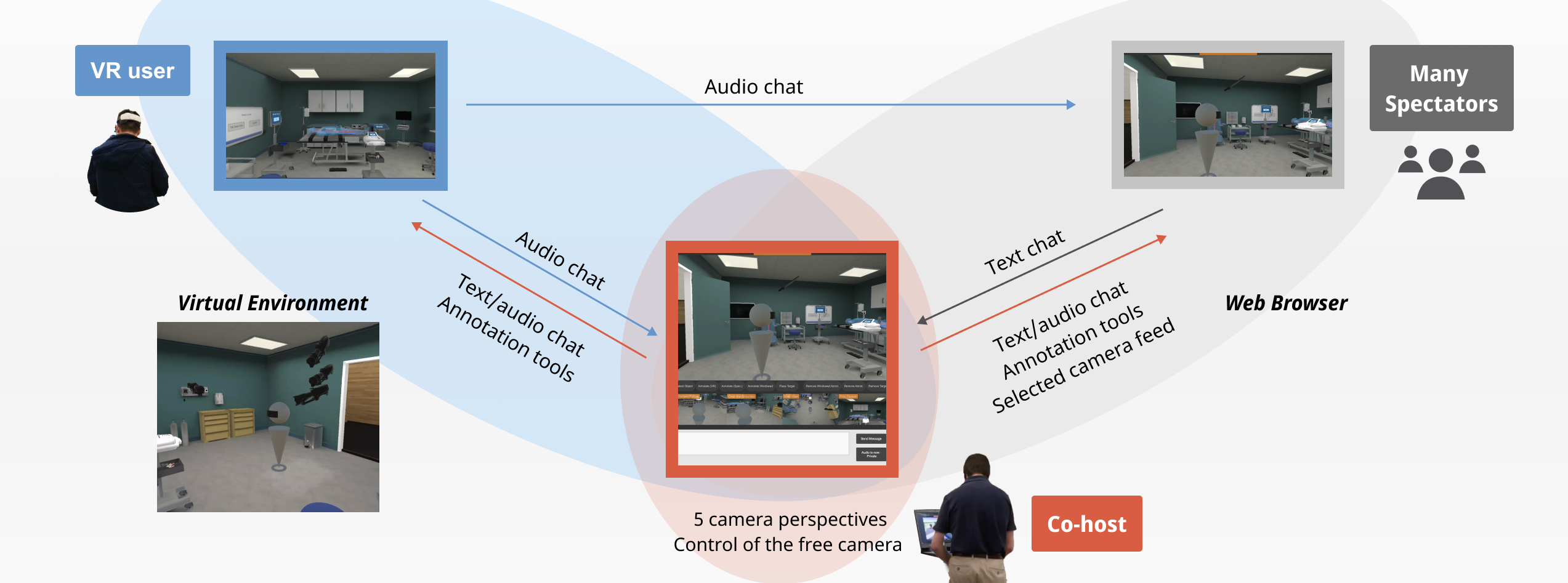}
  \caption{Live stream using \ourtooln; the co-host (middle) is facilitating a live stream of the VR user (left) in the virtual environment through web browsers on laptops and mobiles. Many spectators (right) share the viewport of the co-host.}
  \label{fig:overview}
  \Description{An overview of \ourtool in a virtual scene. The co-host stands in the middle using \ourtooln, with the VR user standing on the left connecting to the system. Spectators view from the same angles as the co-host but using their own devices.}
\end{teaserfigure}

\maketitle

\section{Introduction}
Virtual Reality (VR) is an increasingly popular medium for different applications ranging from art, design, medicine, and entertainment. Like with other media, a user in VR may want to share their VR activity with other spectators while they perform a task. However, these spectators typically view these as 2D video streams on mobile or desktop devices. This asymmetry in interface modality between the spectators and the VR user can create various challenges.

There is a growing body of research\cite{TransceiVR, TutoriVR, DreamStream, XRStudio, Marks} that studies these challenges and proposes systems and interaction techniques to address them. For example, ShareVR \cite{ShareVR} and TransceiVR \cite{TransceiVR} support asymmetric communication to allow a VR user and a non-VR user to collaborate. In situations like remote learning and VR live-streaming, the ability to support multiple views and many passive spectators are critical. XRStudio \cite{XRStudio} achieves the remote multi-user sharing of VR educational content. DreamStream \cite{DreamStream}, and Vreal \cite{vReal} allow multiple spectators to be within the scene of a VR application and view it using conventional desktop interfaces or a VR headset. 

A major caveat of many of these works is that they focus on interactions that only support small groups of users, as defined by Grudin \cite{Grudin}. Interactions used in these works would not work when there are 100s of users. Systems like ShareVR and TransceiVR are limited to dyadic interactions. In DreamStream and Vreal, though the streaming technology is scalable to multiple users, having 100s of avatars in the scene would render the VR application unusable. Additionally, some of these approaches require spectators and non-VR users to install dedicated applications and sometimes access a VR/AR headset.

In contrast, with conventional 2D interfaces, techniques and methods exist to perform interactions at scale, such as chat polls and live sentiment analysis of text chats \cite{EmoteControlled}. Text chats are today's commonly used technique for spectators to interact with a VR user. Such a textual medium can be challenging if spectators need to communicate actionable queries to the VR user. Besides asymmetry, issues also arise due to the number of spectators. It is hard for the VR user to keep track of users' text chats and effectively share their VR activity with everyone. In this paper, we directly address these challenges.

We interviewed six experts with a background in either VR streaming or eSports broadcasting. These formative interviews shed light on the nature of interactions that occur when a VR activity is live-streamed and shared with a large audience. Additionally, it also identifies key challenges in facilitating any form of meaningful interaction between the audience and the VR user. To address these challenges, we propose that an additional user could help. We term this additional user the \textit{co-host} in our work. Their role is to mediate the communication between the VR user and the many spectators. We introduce \ourtool to support the co-host in this mediation. We assume that many users still use text chat to interact. The co-host checks these chats and passes on only the most relevant messages to the VR user. The co-host might also need the ability to direct and control the video stream seen by the spectators. \ourtool allows the co-host to control the viewport that the spectators see. Additionally, it will enable them to annotate the VR scene, place target markers, and highlight arbitrary objects. 

We conducted an informal qualitative evaluation of \ourtool with eight participants who performed the role of a co-host. They used \ourtool to mediate the interaction between a simulated text chat audience, and a confederate VR user. The evaluation sheds light on some of \ourtooln's features in aiding the co-host and identifying shortcomings and scope for future work. Overall, our participants found that using \ourtool as a co-host can add value to the streaming experience for both the VR user and the spectators, as they could better present the scene for the spectators and relay important messages from them to the VR user. 

In summary, the key contributions of this paper are:

\begin{itemize}
    \item Formative interviews that shed light on audience interactions present in the live streams of VR activity and eSports.
    \item \ourtool system, which aids a co-host user to facilitate efficient communication between the VR user and the large audience.
    \item A summative evaluation of \ourtool that focuses on the role of the co-host.
\end{itemize}

\section{Related Work}

\ourtool allows a streaming experience to have a co-host who helps mediate the communication and interactions between a VR streamer and the many spectators who watch the stream. In this, the streamer wears a VR headset. However, the co-host and the spectators do not. Thus this is a form of asymmetric interaction. This interaction can also be seen from the lens of CSCW systems for VR. There are three prior research domains: Multi-user VR systems, Asymmetric interaction in VR and Live streaming.

\subsection{Multi-user VR systems}

Research on multi-user systems in digital 3D virtual environments has been carried out for a long time~\cite{ens2019revisiting, churchill1998collaborative}. Among the earliest systems was the DIVE system~\cite{carlsson1993dive}, in which users are part of a shared 3D virtual environment in which they can share the screens of their 2D desktop applications. \textit{Populated Web}~\cite{www3d} used a multi-user 3D web browsing experience in which web pages are laid out in a shared 3D virtual space. Using a 2D desktop interface or VR headset, users can spatially navigate web content and represent themselves as avatars, interacting with each other through text, voice, and video. Though the work focuses on web browsing, the system's core interactions are strikingly similar to those of today's social 3D virtual environments such as Horizon Workrooms \cite{horizon}, Mozilla Hubs \cite{mozilla}, and VRChat \cite{vrchat}. 

In contrast to spectating in an immersive environment, video streams offer a degraded experience. Video streams are limited in conveying the scale and the richness of the 3D experience of the player. Inspired by multi-user VR systems, prior work has designed and developed techniques that allow the spectators \cite{vReal, DreamStream, JackIn} to be within an immersive environment for spectating the activity of the VR user. However, such approaches do not scale for many spectators. Offering independent viewports to each spectator could lead them to miss out on what the VR user does, which is not desirable in the context of live streams. Furthermore, there is an issue of widespread accessibility where every spectator would want to or have access to a VR headset. Thus, in this work, we focus on retaining the asymmetry and attempt to enhance the interaction where the spectators continue to spectate using video-based interfaces accessible through a web browser.

\subsection{Asymmetric interaction in VR}

Asymmetric interaction between VR and non-VR users has been previously studied in different contexts. The asymmetry is derived from the game design, which provides other visual representations and perspectives to create motivations for users to collaborate \cite{keeptalking}. Another aspect of asymmetry comes from the difference in interactions possible in these mediums \cite{frontiers_immd}. Due to these asymmetries, often, such interactions are challenging because users operate with different visual fidelity and interaction affordance \cite{frontiers_immd}. Sometimes these challenges are intentionally designed. For instance, a widespread use case is collaborative games between VR and PC players who work together to achieve a shared goal \cite{ironwolf,carly,keeptalking,Eyeinthesky}.  

In most cases, these challenges hinder efficient interaction between users. To mitigate these asymmetries in visual fidelity and interaction affordance, prior work has explored how to best share the view of a VR user with the external user. Ishii et al. \cite{Ishii} developed a CAVE-based visualization method that shares the VR experience with bystanders using translucent screens. MagicTorch \cite{MagicTorch} helps communications between a VR user, an AR user, a tablet user, and the physical space. Users can communicate verbally in the co-located mode and through gestures in the online mode. In OmniGlobeVR \cite{OmniGlobeVR}, the first-person view of the VR user is projected on a globe for external collaborators to view the 360-degree video. More recently, ShareVR \cite{ShareVR} projects the VR scene elements on the ground, allowing non-HMD users to interact with the VR user in the same physical space. TransceiVR \cite{TransceiVR} takes an application-agnostic approach, utilizing VR platform APIs to share the VR scene and provide asymmetric collaborative features without source code access. DreamStream \cite{DreamStream} is also application-agnostic and allows a spectator to view a 3D reconstruction of a VR scene. 

All these works require additional custom hardware devices or software on the viewer's end. Most of these approaches and interaction techniques do not scale when the VR user shares and needs to interact with many non-VR users. \ourtool address this issue and provides an interface for a second user (termed as a co-host) to mediate the communication between the VR user and the many non-VR spectators. Unlike systems in prior work, it uses web-based interfaces to make communication cross-platform and scalable, with minimal setup and no additional software for access. Specifically, \ourtool utilizes interaction techniques from TransceiVR to facilitate efficient communication between the co-host and the VR user. However, these interactions enable the co-host to efficiently mediate the communication between many non-VR spectators and the VR user.

\subsection{Live streaming}

Sharing and spectating live streams is a popular online activity today. In such live streams, a video stream of the user's activity is typically captured and streamed to a large audience. Often, spectators typically interact with the streamer and other spectators using real-time text chat and emote reactions. Prior research has identified the different kinds of spectators and how they can influence the streamer's actions \cite{StarCraftFromStands, 10.1145/2465958.2465971, Stahlke_FalloftheFourthWall}. Some spectators may be part of the streaming production, wherein they can directly request a streamer to perform a particular action. Other spectators who are the audience may also request the streamer to perform specific actions with the software and comment and ask questions about the activity. The streamer and the production team could use such audience interactions to adapt the streaming activity in real-time to suit the needs of the broader audience. From the formative interviews we carried out in this work, we found that sometimes such live streams can have a dedicated production team consisting of directors and camera operators, amongst others. 

It can be additionally challenging within the context of live streaming of VR activity. This is due to the asymmetries discussed earlier. The VR user operates in an immersive environment while the spectators view a shaky 2D video stream. Recent research works such as XRStudio \cite{XRStudio} and DreamStream\cite{DreamStream} mitigate this issue by providing dedicated client applications to the spectators that allow every spectator to control their viewpoint. With XR studio, the 3D assets are downloaded to the spectator's client applications. With DreamStream, depth buffers and video frames are transmitted. At the spectator end, these are rendered as 3D point cloud, which allows the spectators the freedom of viewpoint. Both these approaches assume spectators to have sophisticated hardware capable of running 3D applications and providing controls beyond those present in today's video streaming services. 

In our work, we focus primarily on video-based interactions that are carried out on web-browsers that are hardware-agnostic. To this end, today, commercial systems  exist to address this asymmetry specifically for spectating VR live streaming \cite{liv, mixcast}. Instead of providing dedicated applications to the spectators, they offer assistance with enhancing the produced video feeds of the live VR stream. They allow a VR streamer to manipulate the camera viewport of the video streamed to the spectators, and stabilize them. However, these commercial systems do not allow for any meaningful interaction by many spectators with the VR user beyond a simple overlay of the entire text chat. To address this, \ourtool allows a streamer to have an additional moderator (termed as co-host in this work) and aids the latter in mediating the scalable asymmetric interactions between the VR user and the many spectators. It also provides them with controls such as the ability to place 3D overlays and annotations and change the viewport of the video. Our formative interviews with experts revealed that such a separate user who can act as a moderator could add value and facilitate meaningful interaction. 

\section{Formative Interviews}

We conducted formative interviews with experts to better understand the nature of interactions that occur when a VR user shares their activity live with a large-scale audience. In total, we interviewed six experts. Four of them were professional VR steamers whom we will refer to with labels E1-E4. They have 12K, 120K, 380K and 3.3K subscribers correspondingly on their streaming channel, and each have been streaming VR content for the last 6-7 years. We also interviewed two experts (E5-6) who have experience directing the production of non-VR live streams. This helps us understand if similar challenges are present in the live sharing of conventional media. Since these conventional media have been around for longer, we take inspiration from these to inform the design of \ourtooln.

Our experts noted that different members of the audience may have different goals from the live stream. There are broadly three types of people who watch a live stream: (1) people who want to interact with the streamer; (2) people who mainly participate in social interactions with other audience members in the text chat channel; and (3) people who passively spectate the stream without any interaction.

\subsection{Challenges}
The interviews elicited challenges associated with live-streaming VR content to a large audience. A significant issue mentioned by all the experts was the asymmetrical nature in which the VR user and the audience see the virtual environment. While the VR user sees a virtual environment immersively in 3D, the spectators are restricted to seeing a shaky 2D video feed. This creates a barrier to efficient communication and interaction. Such issues have been the extensive focus of some prior works \cite{TransceiVR, TutoriVR, DreamStream, ShareVR} that focus on mitigating challenges in asymmetric interactions. Hence, we skip their discussion and focus on issues that emerge in scenarios where a large audience is present.

\subsubsection{Scalability of text messages}

From our interviews, a key challenge noted by our experts was a streamer's ability to engage effectively with their spectators, especially in the case of a large audience. Spectators' only mode of interaction with the streamer is through text chat. Today many streamers use widgets that overlay these chats inside existing VR applications. In the case of a larger audience, messages can be intended for fellow spectators and not the streamer, and also those that are generic and non-actionable. 

E1: ``\textit{When I was a smaller streamer, if you're going to be talking with two or three people, that's fine. The bigger you get, the more confusing that dialogue becomes. I would even argue if you get to five people, now a conversation begins to happen just between those people.}''

To this end, our experts noted different techniques that help them to have an effective interaction. E1 built a custom Text-To-Speech chatbot that would convert any message prefixed with a specific character to an audio message for the streamer to hear. This allows the spectators to allow for channeling specific messages to an audio channel that the VR user can listen to, instead of having to parse through all the messages. However, this relies on spectator etiquette and could fail when multiple spectators want to interact simultaneously. E3 confirmed this issue and alluded to the need for having a separate moderator for the text-to-speech conversation. Along those lines, when E2 streams VR activity, they have another person alongside, who looks at the desktop window, reads, and relays the important chat messages to them.

\subsubsection{Interaction Latency}

Our experts noted that there is a latency associated with the video stream that the spectators view, and it is usually of the order of 20s to a minute. Such latencies are typical of systems that need to encode and distribute video to many viewers over the internet. This means that sometimes a chat message sent by a spectator could refer to an action or scene element from the past, and it might be too late by the time a VR user sees it. This hinders effective interaction and could potentially be confusing to the VR user. Our experts speculated that it could be interesting to explore interactions beyond text chat, such as providing the spectators the ability to interact directly with the game or trigger something. However, they noted that the latency issue could render such scene interactions unusable.

\subsubsection{Choice of perspective}

Our experts had varied opinions on which perspectives could be more useful. Some of the experts felt that the First-Person View (FPV) feed of a VR user is a more faithful representation of the VR user's experience and tends to be more engaging. However, such FPV feeds can be more shaky and uncomfortable to watch. This is especially prominent in scenarios when the VR user moves their head a lot, such as in action sequences. To address this, all our VR experts mentioned that, they intentionally slow down their head movements, actions and sometimes even exaggerate their hand actions so that viewers can follow this. Additionally, they make sure to not talk about something in the scene, that may not be in the FPV feed. Such issues were also found in prior work \cite{TransceiVR, TutoriVR}.

In contrast, a Third-Person View (TPV) is more stable, and could allow a spectator to get a better perspective of the scene. Such varying preferences are also corroborated by prior work \cite{StreamingVRPerspectives} that surveyed the preferences from viewers. However, this means that either the TPV needs to be chosen strategically depending on the VR activity; alternatively, the TPV can be changed dynamically by the VR user but this could add to additional cognitive load for them and was noted by E2. E1, E5 and E6 noted that, even within the realm of a TPV camera, the choice of TPV, such as an elevated map view, or a view that follows a player, is informed by the type of application as well as the context of action in these applications. Such streaming view control tools are not popular for VR applications, and the ones that exist are harder to use to achieve real-time camera control. However, for systems that stream conventional media, E5 and E6 note that these systems have well-understood mechanisms and systems for this. They tend to have specific templates for views e.g. FPV, views that focus on player's facial expression or action, and have quick shortcuts to transition the camera to these pre-defined template views. There's a dedicated person or team that controls these cameras in real-time. Sometimes, the live audience text chat may influence the real-time choice of the camera views that is streamed.

\subsection{Design Goals}
In summary, our formative interviews suggested that an additional person, equipped with appropriate tools, could help address the challenges for VR users of interacting with large audiences. Such a tool should have the following characteristics for this {\em co-host}:
\begin{itemize}
    \item Allow for filtering and selective relaying of text chat messages via private text or audio
    \item Support low-latency interaction between co-host and VR user
    \item Enable flexible camera control independent of the VR user.
\end{itemize}

\section{The \ourtool system}

\begin{figure*}[h]
  \centering
  \includegraphics[width=\textwidth]{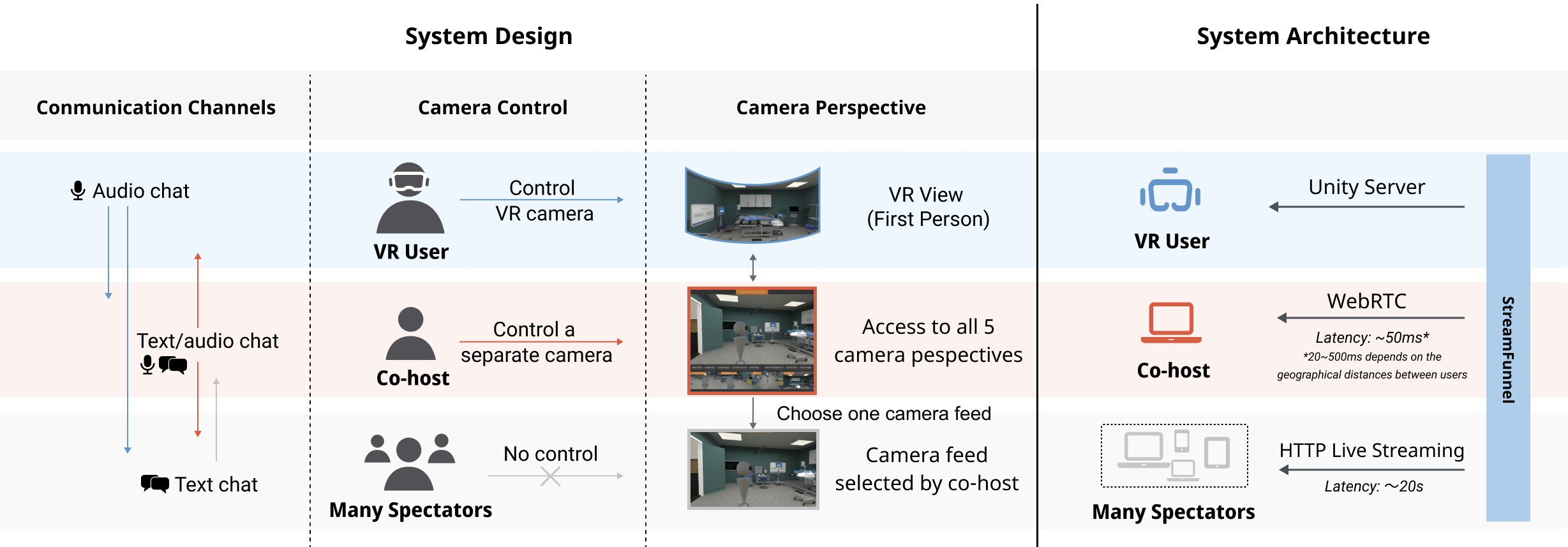}
  \caption{Three-tiered interaction system with VR host, the co-host, and spectators}
  \label{fig:system}
  \Description{Left panel shows the three-tiered system design: VR user controlling VR view, the web co-host controlling the streaming camera, and spectators viewing the streamed view. Spectators have no control over the camera. Right panel shows the system architecture, where \ourtool connects the VR user through Unity Server, the web co-host through WebRTC, and the spectators through HTTP Live Streaming (HLS).}
\end{figure*}

To implement the design goals identified in the formative interviews, we present the \ourtool system. As shown in Figure \ref{fig:system}, \ourtool is a three-tiered system with (1) a VR user running a Unity scene containing the \ourtool utilities, (2) a co-host connecting remotely through WebRTC, with their own scene camera and various interactive features, and (3) spectators who observe the scene through HTTP Live Streaming (HLS) and Amazon Web Services (AWS) \cite{aws}. While the co-host enjoys a high degree of freedom and control in the scene, the spectator role resembles viewers of traditional live-streaming; they can only view the video streams rendered from the co-host's view. 

\begin{figure*}[h]
  \centering
  \includegraphics[width=0.9\textwidth]{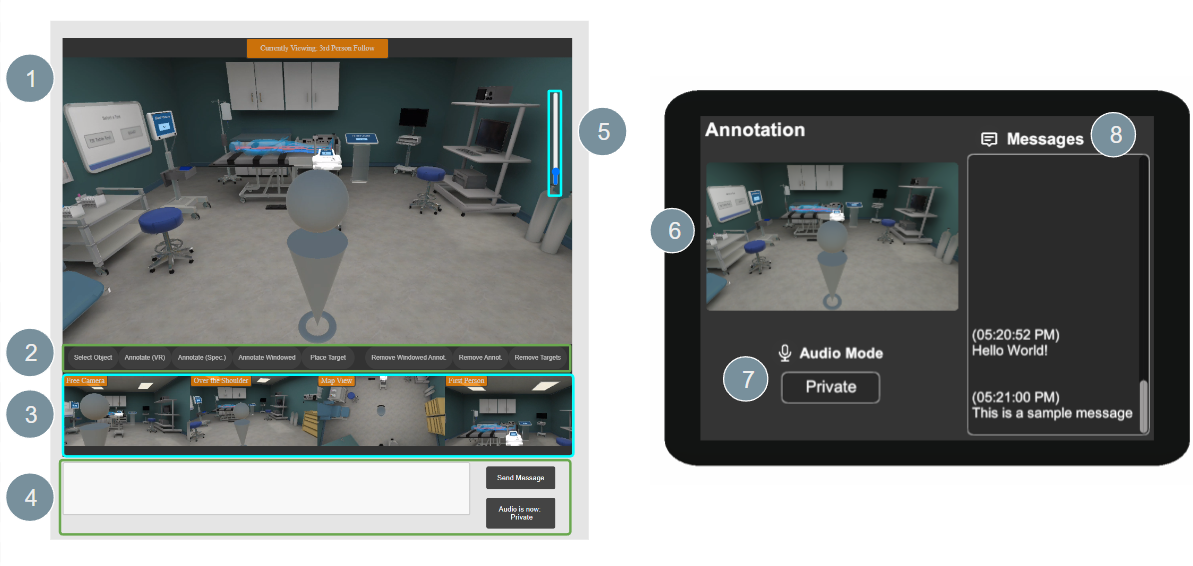}
  \caption{User Interface for the web co-host and for the VR tablet. (1) - The captured view of the camera in focus, (2) - a toolbar of asymmetric interactive tools, (3) - four other camera feeds from preset cameras, (4) - private text-audio controls, (5) - a slider to adjust camera angles, (6) - the snapshot of streaming content on the VR user's tablet, (7) - VR user's audio indicator, (8) - messages from the co-host.}
  \label{fig:ui_interface}
  \Description{Left panel shows the user interface of the web co-host, where a view of the medical room (Task B) is overlaid by a camera label at the top, and a toolbar at the bottom. The toolbar includes buttons for: select object, annotate (VR), annotate (Spec.), annotate windowed, place target, remove windowed annotations, remove all annotations, and remove targets. Under the toolbar, there are four smaller camera feeds available for the co-host. At the bottom is a textbox along with one button for sending messages and one button to indicate the VR user of the current audio state (private/public). Right panel shows the user interface of the VR user (the VR tablet), where a view of the same camera in focus is shown. The tablet also includes the audio state indicator (public/private) and message history.}
\end{figure*}

\ourtool provides three groups of functionalities to the co-host, including camera motion presets, asymmetric interactive features through web browser, and private text-audio controls. The user interfaces for the web co-host and the VR user are shown in Fig. \ref{fig:ui_interface}.

Many of \ourtooln's design elements are motivated by challenges raised by experts in the formative interviews. Discussions on handling excess chat messages prompt us to introduce the co-host role, who will filter useful information with the assistance of private text-audio controls. Since all roles are remote, we also bring latency into \ourtooln's design consideration. In the system's network architecture, we use the WebRTC framework to provide real-time, highly interactive participation for the co-host, who would typically experience a 20-500 ms latency depending on the geographical distance between the two users. This allows the co-host to work seamlessly in the scene with the VR user, making solutions to asymmetric interactions possible. On the other hand, HLS through AWS could provide a traditional live-streaming experience that can be accessed by potentially thousands of spectators, allowing general scalability of access. For concerns related to perspectives, \ourtool provides various camera motion presets to the co-host, allowing them to choose FPV, TPV, or other options depending on the context. 

\subsection{Camera Motion Presets}

Access to multiple camera perspectives is common in program production and commentary. As discussed in the formative interviews, in the VR setting, perspectives beyond the first-person view could be beneficial in providing stable camera feeds and transient context information. However, relying on the VR user to dynamically change camera feed during the VR activity can be cognitively overwhelming. Hence, \ourtool addresses this by providing the access to multiple camera motion presets for the co-host. The co-host can switch between views and select which view will be presented to the spectator, which is similar to the role of directors in traditional television studios or commentators in eSports commentary. 

\begin{figure*}[h]
  \centering
  \includegraphics[width=0.9\textwidth]{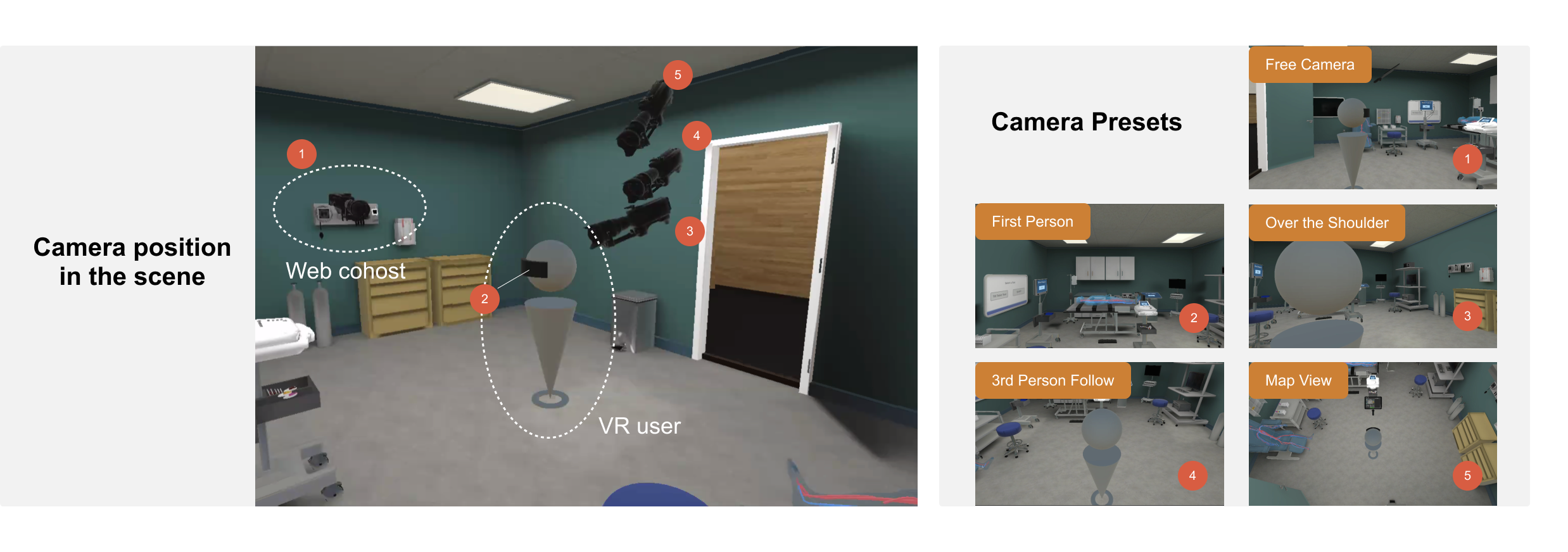}
  \caption{Camera presets for co-host stream}
  \label{fig:camera_presets}
  \Description{Left panel shows an avatar standing in a testing room, with five cameras around it at different angles. Right panel shows the captured views of the five camera presets the co-host can select: free camera, first person, over the shoulder, 3rd person follow, and map view.}
\end{figure*}

As shown in Fig. \ref{fig:camera_presets}, \ourtool provides five cameras for the co-host, with one main camera and four preset cameras. The co-host can move the main camera freely inside the VR user's environment using keyboard and mouse controls, giving the co-host a sense of co-presence in the virtual scene. The four preset cameras include 3rd-person follow, over-the-shoulder, bird's-eye (map), and first-person views. These preset cameras would automatically follow the VR user as the user moves or looks away; they provide important contextual information around the VR user. Once switched to the 3rd-person follow, over-the-shoulder, or bird's-eye view, the co-host could also change the virtual camera arm length using a slider, as shown in Fig. \ref{fig:ui_interface}.5. This mimics how real camera operators adjust the camera positions in studio productions.

To allow better management of the virtual scene, we design features for the VR user to adjust the position of the co-host's camera. The VR user can grab the main camera and move or rotate them to a desired position, hence changing the co-host's viewpoint. This comes in handy for tutorial purposes, where the VR user wants to display static angles or perspectives of their actions, as it is hard to communicate camera positions and directions over audio. We also provide the VR user with access to a virtual tablet (Fig. \ref{fig:ui_interface}) that allows them to examine the current camera feed spectators are viewing.

\subsection{Web-based Asymmetric Interaction}

\ourtool helps mitigate the interaction barriers that arise due to the asymmetrical nature of interaction between the co-host and the VR user. It takes inspiration from the interactions introduced by TransceiVR \cite{TransceiVR}, builds on them and makes them available to the co-host through a web browser accessible remotely. These are showcased in Fig. \ref{fig:tools_diagram}, and we describe them below.

\subsubsection{Scene Annotation} (Fig. \ref{fig:tools_diagram}.a) A scene annotation function allows for the non-VR co-host to draw an annotation over any part of the scene or objects, which is adjusted to be placed at the appropriate depth. The co-host sends the mouse coordinates to the Unity app, which maps them to 3D coordinates based on the depth of the objects in the scene raycasted from the camera. \ourtool offers two annotation options -- VR and Spectator mode -- to the co-host; for each mode, only the VR user or the spectators will be able to see the annotation, respectively. This provides the co-host with control over who should see the information.

\subsubsection{Windowed Annotation} (Fig. \ref{fig:tools_diagram}.b) A second type of annotation function is a windowed annotation.  Instead of rendering a depth-corrected annotation object within the scene, the windowed annotation feature allows the co-host to draw an annotation rendered on top of a snapshot of the spectator's view that is only displayed on the VR user’s tablet. This allows the co-host to convey information to the VR user from the spectator's perspective.

\subsubsection{Place Target} (Fig. \ref{fig:tools_diagram}.c) The co-host can place a blue target object in the scene which is visible to all other spectators and the VR user. The co-host clicks on the scene viewport to place the target on desired surfaces. This allows guidance on directions or more exact locations.

\subsubsection{Select Object} 
\begin{figure*}[h]
  \centering
  \includegraphics[width=\textwidth]{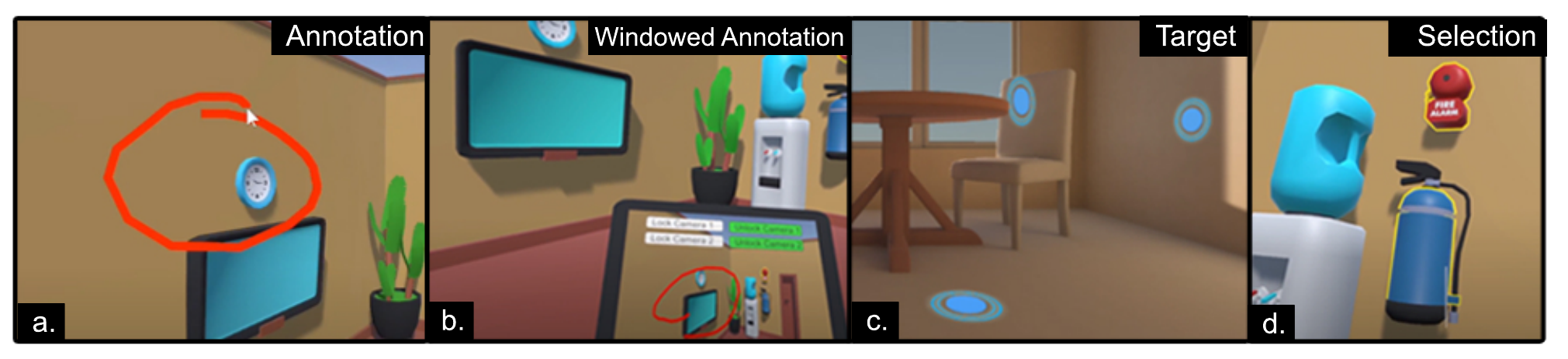}
  \caption{Asymmetric Interactive features for the co-host}
  \label{fig:tools_diagram}
  \Description{First panel shows the annotation feature, where the web co-host draws a red circle over a clock in the scene. Second panel shows the windowed annotation feature, where the web co-host draws a red circle over the clock, and the circle is only shown through the VR user's tablet. Third panel shows the target feature, where the web co-host places several blue circle targets on walls and furniture surfaces. Last panel shows the selection feature, where the web co-host highlights a fire extinguisher and a fire alarm on the wall, and a yellow outline surrounds their 3D meshes.}
\end{figure*}

(Fig. \ref{fig:tools_diagram}.d) This feature allows the co-host to select objects in the Unity scene utilizing raycast functionality. Clicking an object places an outline over the 3D mesh of the object in the scene, helping the co-host to draw attention on a particular object to the VR user or the spectators. The co-host can select and deselect multiple objects within the scene.

\subsection{Private Text-audio Controls}

As in many collaborative production scenarios, performers and the staff will need private communication channels. \ourtool provides both text and audio controls aiming to assist such communication. 

As discussed in the formative interview, it can be challenging for the VR user to deal with flowing chat messages from a large audience in immersive environments. The co-host hence bears the duty of filtering spectator chats in \ourtooln's design. We combine the common practice of text overlay with moderation from the co-host as suggested by our experts (E1, E2). The co-host can choose to relay the selected information and questions they filtered through private audio or the text overlay visible on the VR user's tablet, which will stay available on the tablet throughout the session. In scenarios where the VR user needs to pay full attention to the tasks at hand, this could help retain important feedback from the spectators. \ourtool also includes an audio state indicator, much similar to the ``On Air'' sign in television or radio studios. The co-host could turn it on or off to tell the VR user if their audio is being broadcast to the spectators, allowing private channel communication through audio. 

\subsection{Network Architecture}

A cross-platform, robust networking solution is required to support \ourtooln's three-tiered system with a VR user, a low-latency co-host, and hundreds to thousands of spectators. WebRTC (Web Real-Time Communication) \cite{webrtc} is a real-time peer-to-peer connection framework that has been explored in many video conferencing and educational scenarios. Prior systems include USE Together \cite{USETogether}, Gunkel et al.’s 360-degree social VR experience \cite{Gunkel}, XRDirector \cite{XRDirector}, and more recently, XRStudio \cite{XRStudio}. \ourtool takes inspiration from these prior works and expands on them. We use Unity and its Render Streaming functionality to integrate WebRTC with existing applications and achieve better synchronous collaboration. Unity also allows us to increase rendering rate easily, which could efficiently reduce server-side latency for a WebRTC-based cloud XR system as pointed out by Vikberg \cite{Vikberg}.

\begin{figure}[h]
  \centering
  \includegraphics[width=\columnwidth]{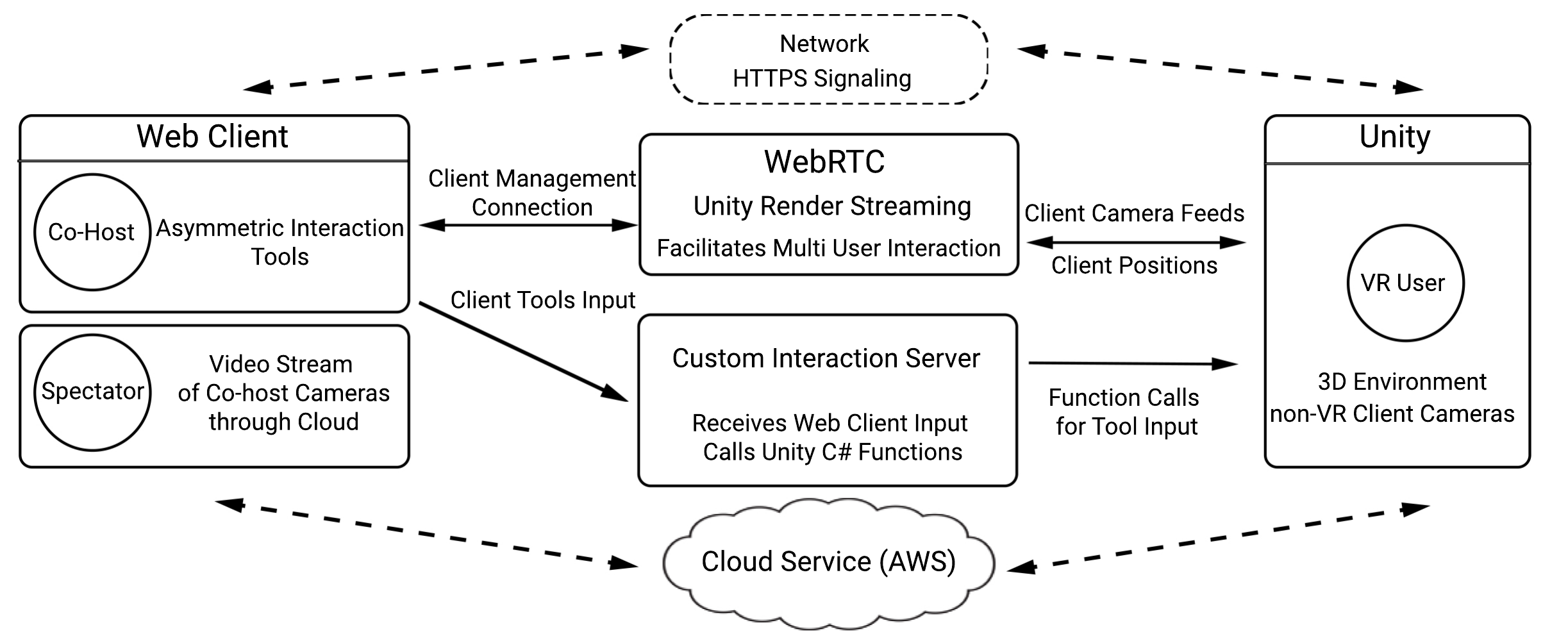}
  \caption{The underlying network architecture of \ourtool (STUN and TURN servers not shown)}
  \label{fig:websystem}
  \Description{Three components: web clients, \ourtooln, and Unity. Web clients include the co-host and spectators. \ourtool includes HTTPS signaling, WebRTC, custom interaction server, and cloud service. The co-host connects to Unity through HTTPS signaling, WebRTC (video feed), and custom interaction server (tool input). Spectators connect to Unity through cloud service (AWS).}
\end{figure}

\subsubsection{WebRTC and Unity Render Streaming}

We design a WebRTC-based network architecture to support communication between the Unity VR application and the web front, as shown in Fig. \ref{fig:websystem}. Devices connecting as the co-host to the Unity server on the VR user's computer establish a WebRTC peer-to-peer connection with the computer through the Unity Render Streaming package. The peer connection is established with help from three servers: a signaling server, a public STUN (Session Traversal Utilities for NAT) server that discovers IP addresses for devices, and a TURN (Traversal Using Relays around NAT) server that relays traffic when direct connection fails. A signaling server, deployed on AWS, serves as the central relay server that handles the initiation of the peer connection. The signaling server will relay offers from the server to the co-host client and vice versa; once offers and answers are exchanged, multimedia channels will be created for the cameras, streaming the rendered video from the VR program to the web frontend. A TURN server we deployed on AWS will automatically relay the streaming traffic if the direct peer connection becomes unstable.

\subsubsection{Custom HTTP Interaction Server and WebSocket Relay Service}

The co-host is provided with an array of interaction tools with which they can interact or mark up objects in the Unity scene. The actions performed by the co-host client using these interactive features are relayed to the Unity machine by sending JavaScript GET/POST requests to a custom interaction server running inside the Unity application. The server implements a REST API and parses the requests from the client, which includes identifying information along with the appropriate parameters necessary to perform specific actions (such as mouse cursor position). To help with the complications of NAT and firewalls during connection, we implemented a relay service on the central signaling server utilizing the WebSocket protocol. This allows us to open up bidirectional communication through TCP connections, relaying requests and responses between the web client and the Unity application.

\subsubsection{Spectator Connection}

Spectators connect to \ourtool using a different URL than devices wishing to connect as the co-host. The rendered output streams of the co-host's in-focus camera will be broadcast from the VR user's machine to the AWS Elemental Live service using FFmpeg \cite{awslive, ffmpeg}. The service then buffers and transcodes the video streams into multiple video resolutions; connecting spectators can then view those video streams from the AWS storage using HTTP Live Streaming (HLS). Building the spectator component on the cloud allows the VR user to scale the service on demand, allowing a very large group of spectators.

\section{User Study}

To understand the benefits and shortcomings of the \ourtool system, we conducted two sets of informal evaluations: one expert evaluation with two experts and a user study with eight participants. The expert evaluation focuses on how the system responds to streamers' needs, while the user study focuses on the co-host's experience using the system.

\subsection{Expert Evaluation}

We recruited two professional VR streamers as experts (S1-S2) in our study; they have 4K and 107K subscribers in their channels, both with streaming experiences of three to four years. Note that they are different from experts E1-E4 in the formative interview. We interacted with them for 80 minutes, and the experts were compensated for their time. During the evaluation, the expert watched a brief introduction video on the system design; we demoed the system with a VR game resembling a typical VR live stream. In a semi-structured interview, we asked the experts if they feel the system responds to the challenges mentioned in earlier sections from the streamer's perspective and other general feedback. Their responses are discussed together with results from the user study.

\subsection{User Study}

We also conducted a user study to investigate the co-host's experience facilitating communication between the VR user and the spectators. We recruited eight participants (P1-P8) (Five male, three female, age range 18-27). All of them had prior experience with VR, but not necessarily with live-streaming or teaching experiences. Each study lasted 75 minutes, and the participants were compensated with a \$35 gift card for their time. An author of the paper took the role of the VR user and hosted the VR part of the system remotely. Participants used \ourtool through Chrome on a WiFi-connected laptop away from the remote VR player. While both joined the system over the Internet, they were also in a Zoom meeting where they could hear each other speaking. Another author of the paper was also in the Zoom meeting, in which they simulated remote spectators and sent pre-generated chat messages to the zoom chat window. Before the study, all participants were given a 20-minute training session, where they watched a tutorial video and went through a virtual training scene using the \ourtool system. Participants then performed the role of co-host in two tasks (Task A \& B), in which the order of the tasks was counterbalanced. Each task took 10-15 minutes and was followed by a post-study questionnaire. An author of the paper conducted semi-structured interviews with the participants after they finished both tasks. The following sections discuss the task designs, measures, and results. 

\subsection{Task Designs}

\begin{figure}[h]
  \centering
  \includegraphics[width=\columnwidth]{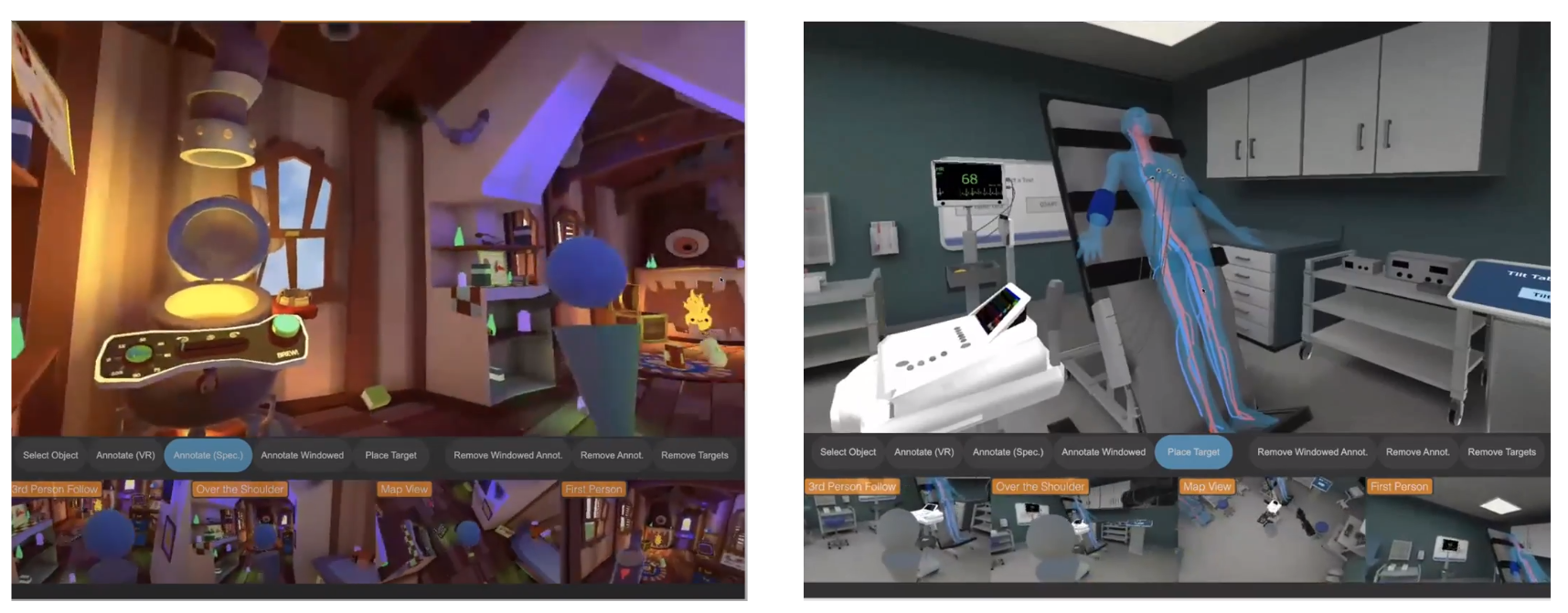}
  \caption{Escape Room game (left); Medical instructional room (right).}
  \label{fig:scene}
  \Description{Left panel shows the VR user standing in an escape room scene. Right panel shows a medical testing room with equipment.}
\end{figure}

We design two tasks for the user study, one resembling a VR game live-streaming (Task A) and another as an online VR lecture (Task B) (Fig. \ref{fig:scene}). The participant performs the role of the co-host, who has a two-part goal: they need to assist the VR user in presenting the experience and relaying the spectators' requests to the VR user. They are asked to choose features from the tools we provide that they find most helpful in achieving these goals, including camera motion presets, asymmetric interactive features, and private text messaging. The VR user cannot see the entire chat messages from spectators, and the simulated spectator can only see the camera feed chosen by the co-host. 

\subsubsection{Task A: VR Game streaming}
In this task, we simulate a VR gaming live stream where the VR user is playing a VR escape room game accompanied by the co-host and streamed to the spectators using \ourtooln; the game is modified from Unity's VR Beginner tutorial \cite{VRBeginner}. The VR user needs to first find instructions and information sheets around the starting position and then move to the other side of the room. The VR user will need to interact with objects in the scene (wand, cauldron, and ingredients) to make a potion, which is used to open the door and exit the room. The co-host is expected to present an entertaining view of the game experience and relay valuable comments from the chat messages.

\subsubsection{Task B: Online VR Lecture}
In this task, we simulate an online lecture where an instructor (the VR user) and a teaching assistant (the co-host) demonstrate how to conduct a medical testing procedure. The VR user performs the test by interacting with the scene objects, such as a blood pressure cuff and ECG electrodes and narrating necessary information. The co-host is expected to position the spectator's view to optimal locations and relays questions or information from the spectators to the VR user as they are busy performing the procedure. The VR user will also occasionally asks the co-host to highlight objects in the scene or focus the view on certain locations.

\vspace{10pt}
The two scenarios simulated in the tasks share some common features, where (1) they typically involve hundreds to thousands of spectators in real life, (2) they require the VR user to concentrate on missions inside the virtual environment, and (3) communication between the VR user and spectators is crucial in both scenarios. They differ in content and expectation. Generally, a game stream like Task A is more fast-paced, as the VR user constantly moves and engages with objects in the game scene; in lectures like Task B, the VR user will occasionally pause their movement for narration. Comments will show up in greater volume and at a higher speed in Task A compared to those in Task B. There is also a greater percentage of non-actionable and casual comments in the stream, while comments in the lecture are mostly topic-related questions and feedback. Such differences create greater pressure on the co-host in Task A, requiring them to pass information more efficiently. Assumptions on the visibility of the co-host to spectators are also different. The VR user talks to the co-host and directs them for specific actions during Task B, while in Task A the VR user only speaks to the spectators.

\subsection{Measures}

Participants of the study completed questionnaires regarding the usefulness of the different features of \ourtool in presenting spectators with the VR user’s activity and communicating the information from spectators to the VR user. The ratings were based on a 5-point Likert scale. Following the questionnaires, we then conducted semi-structured interviews to better understand the usability of \ourtool in facilitating large-scale interaction between a VR user and many spectators; this also includes open-ended feedback on the system and general study experience.

\subsection{Results and Discussions}

We gathered qualitative feedback from the user study interviews and questionnaires. Overall, all participants agreed that using \ourtool by a co-host can add value to the streaming experience for both the VR user and the spectators. Seven out of the eight participants agreed or strongly agreed that the system could better present the scene to the spectators using features like camera presets and asymmetric interactions. When communicating the spectator comments to the VR user, seven participants agreed or strongly agreed that \ourtooln's private text-audio controls helped relay important information. 

\subsubsection{Perspective Selection of the Streamed Video}

\ourtool provides multiple camera perspectives that allow the co-host to curate the content for spectators and help them better understand the context. The additional camera perspectives also allow a better presentation of the stream and reduce the cognitive load on the VR user, as they no longer need to intentionally pause actions to achieve a stabilized view. In the expert evaluation, both experts expressed positive feedback on the camera perspectives. S1 mentions that \emph{if somebody wanted to re-project their VR live streams for somewhere, sometimes you can't because the footage is awful from the first-person view. But if it's nice and cinematic and has different angles, it could be really interesting.}

Our user study shows that participants were able to utilize different camera perspectives to communicate spatial information and context in the scene; they found that multiple camera views and presets helped showcase elements ignored in a single FPV or TPV perspective. For example, both P1 and P4 used the free camera to address spectators' requests outside the VR user's field of view. Some of them made perspective choices according to the circumstances, as P1 summarized: \emph{"The free camera allows independent focus on subjects of interest that the chat may want to deal with. On the other hand, if it's very hands-on, like lectures or the actual mixing of the ingredients in the game, first-person works fine." } 

P6 chose the third person view camera as a way to engage the audience: \emph{"If I choose the third person view, I can actually be more focused on what the VR user is talking to. I felt that I'm playing with the user and I'm not the only one in this virtual [scene]. And I think the spectators will feel the same way."} 

These participants' decisions demonstrated how the co-host can better tailor to the spectators' needs, when the VR user would not be able to show due to the VR user's limited attention and immersive FPV. This can bring higher engagement from the spectators while communicating the VR user's actions across spectators more effectively.

Participants also paid attention to the camera's locations in the scene. Some placed the camera in certain locations ahead of time to highlight important moments; in Task A (game), P5 placed the free camera outside the escape room and used that as a preset to capture the moment when the VR user successfully exited the room. A few participants cited difficulties navigating to ideal locations using the free camera controls, where positioning the camera takes extra time and could deviate from focusing on the VR user's actions. P5 suggested that it would be helpful to have preset positions in addition to angles, akin to a TV broadcast setup, to mitigate the issues with navigating the main cameras in the scene.

\subsubsection{Referencing Scene Elements to the Spectators}

The goal of annotation and place target features is to support the co-host in referencing objects based on their needs and different contexts. In the expert evaluation, S1 and S2 had different opinions on how engaged the co-host should be when referencing items. S1 found that these features could be helpful in addressing the interaction latency challenge, as the co-host could digest the delayed comments and interact with the VR user with a smaller latency. S2 preferred the co-host to have a smaller degree of engagement in the virtual environment, as the spectators may see an added layer of communication between them and the streamer.

In the study, participants also had varied opinions on how the asymmetric interactive features helped with presenting the scene to the spectators. In the interview, P2 rated annotation and target as the top two useful features that help with facilitation, because \emph{“it’s very natural to draw circles and say, this is what I'm talking about. The second useful is the target because you can do precise pointing instead of very big drawing.”} 

Some participants prioritized timeliness in a fast-paced live streaming, hence they preferred placing a target when they feel the need to give quick response. P6: \emph{“For the game experience play, putting a dot on the object is easier for me than to annotate something. Because I felt that I am running out of time, and I need to succeed, something that I to help the user to complete their goal, so I need something very quick. Pointing out tools may be helpful for communicating with both spectators and users.”} The annotation features which take longer to render are therefore less preferable by some participants. P7:\emph{"[Annotation] takes too long to circle things and then render [...] so just placing a target was pretty useful."}

The visibility of annotations and target to spectators is another consideration of participants when choosing which feature to use. Some participants like P1 choose to use annotation instead of select object in Task B (medical) because annotation is more simple and obvious to the spectators, P6 also used annotation more in Task B (medical) because this participant thought the annotation is much more recognizable for spectators than placing a light blue target on the white game object in the scene.

Participants also utilized the annotation tool to address spectators' questions and reduce the cognitive load of the VR user. P7: \emph{“If the questions were asking to point something out, then it could be as simple as placing a target or changing the view of the camera.”} 

Some participants used the features as an aid to communicate with the VR user. However, uses of these features to interact with the VR user have been studied in prior works like TransceiVR\cite{TransceiVR}; hence we skip its discussion in this work.

\subsubsection{Communicating the Spectator Comments to the VR User}

In the formative interviews, our experts highlighted the issue of scalability, where excess chat messages are difficult for the VR user to process. To address this, the additional co-host role in \ourtool combines moderation and text overlay to reduce the cognitive load of the VR user and better facilitate information. In the expert evaluation, experts shared varied opinions on the moderation of chat content. Both experts agreed that control over chat messages is necessary, but S2 preferred a full chat history to be displayed, where messages that the co-host chooses are highlighted, as some spectators may want their messages to be seen on the screen content.

In our user study, all participants chose to filter a significant portion of the simulated comments for the VR user, where they found it necessary and beneficial when the spectators are at scale. P2 mentioned \emph{"I would definitely do some filtering on my own of what is a good question, what is not a good question. For example, if you're doing game streaming, there's always a lot of trash on the chats, and you do have to filter a lot. So I think a co-host would definitely be a very helpful thing when there are a lot of comments in the chat."}  When asked to elaborate on their decision, participants' rationales include whether the comment is actionable for the VR user or whether the questions are already answered. P7 said,\emph{"there was a question about ‘where did we place the blood pressure cuff?’ The VR user ended up showing that, so I didn't bother to answer that out loud or send it to the presenter."} 

After filtering comments, \ourtool provides both private text and audio controls for the co-host to choose when relaying information to the VR user. Participants made varied decisions on whether to use text or audio chat; overall, they found both channels are viable depending on the situation. Some participants used the audio chat for more urgent communication, such as messages related to transient elements in the scene (P1, P6). Other participants perceive the text messaging feature as a non-intrusive communication method, especially when they think the VR user needs to focus on the current task and should not be disturbed by the chat. We observed this in both tasks. P7: \emph{"when the presenter was talking, I don't want to break the flow of presenting with the questions of the spectators";} P8: \emph{"I figured I'd just give more subtle questions on the chat instead because I didn't want to break the train of thought."} P1 evaluated the current cognitive load of the VR user and decided whether to send the message or use audio:\emph{"in certain cases, it's probably better to leave it and tell the user to take a look at chat when it's necessary to avoid bogging them down with information overload."} 

Participants also considered the needs of spectators in different contexts when determining the timing and method for sending the spectator comments. For example, P5 thought the co-host should have less presence in Task A (game) so that the spectators' focus is on the streamer: \emph{"I just wanted to be the one that's facilitating the communication. And instead of me talking, I wanted to take the audience's responses and deliver that, which is why I try to use the text."} Similarly, P8 also chose to send text messages because \emph{"it felt more disruptive to speak out loud because everyone was there to watch the person who is leading."} In Task B (medical), P8 used audio chat more because \emph{"for teaching assistance, it's more like you're a secondary teacher, and you're speaking to the students and having a relationship with the students."} Such decisions shed light on how an experienced co-host could add value to the spectator's experience while keeping the communication between the VR user and spectators largely unaffected. 

\subsubsection{Impact of Learning Curve and Workload on the Co-host}

During the interviews, we asked participants about the challenges they encountered when using \ourtooln. Some participants mentioned navigation troubles and unfamiliarity with the camera controls may take extra mental bandwidth which could have been used to communicate the chat. P8: \textit{" I was kind of forgetting about the communication aspect because I was just busy trying to figure out how to get around."}

This issue could be mitigated by practicing using the tool, as suggested by participants in our study who feel more familiar with the tool in the second task. For example, P6 who have no experience in watching live streaming found it difficult at the beginning of the first task to figure out what to do, but was able to utilize more features for communication in the second task.

Participants also expressed concerns about not being able to address the concerns of all the spectators on a large scale. P1: \textit{"once you've hit a certain point, the amount of people who want different things at once is going to be very hard to process."}

\section{Limitations and Future Work}

A key task of the co-host is to filter and relay important chat messages to the VR user. Currently, the spectators' chat box is separated from the \ourtooln's communication interface between the co-host and the VR user. The co-host needs to either copy-paste the important comment from chat into \ourtooln, or read it to the user while handling other tasks at hand, making it hard for the co-host to timely manage the information. This might be addressed by providing a more advanced chat interface that supports the co-host in relaying chat messages. For example, integrating chat into the co-host interface and allowing the co-host to quickly tag a text message, which will be automatically forwarded to the VR user. Such support will save the co-host's time in sending out the chat message, adding to the efficiency of communication, especially when there is a large group of spectators. Regarding concerns about the co-host's workload and mental bandwidth, some additional features may include natural language processing systems that can automatically suggest useful chats to the co-host.

The moderation of chat messages also involves a social aspect, where our experts point out that spectators may want to see evidence of their comments delivered to the streamer. Future work on the system could look at different models of moderation that preserve such social interactions.

Current free camera locomotion is constrained to WASD and mouse controls, which can be challenging for the user in navigating the scene. This limitation was mentioned by many participants of our user study as a major challenge. The frustration or failure in controlling the camera could influence the spectators' experience as well as increase the co-host's workload. To address this issue, future work can explore more advanced camera control, including automated and stabilized camera movement, customized camera presets, etc. The system can also provide "best-practice" suggestions for co-hosts who are not familiar with live-streaming, such as recommended camera placement and camera feed selection.

While \ourtool provides the co-host with control over the free camera and different camera perspectives, it is still difficult for the co-host to stay aware of the VR user's position. This could make both the co-host and spectators feel disoriented and lose track of the VR user's activity. Therefore, future work can investigate how to enhance the co-host's awareness of the VR user, such as indicating the VR user's position when the VR user is not in the scene.

Apart from the camera control, in the user study, the spectators can only participate via text chat, and we have not yet investigated non-textual ways of communication in \ourtooln. An avenue for future work is to investigate spatial interactions that can engage large groups of spectators for asymmetric communication. Interaction techniques introduced by prior work, such as Spatial Dynamic Voting \cite{sdv}, Density Maps \cite{10.1145/2168556.2168560, 10.1007/978-3-030-85613-7_38}, and Interactive spatial visualizations \cite{Denning2011}, could be extended to the context of asymmetric communication.
 
A limitation of our findings in the evaluation is that responses were simulated to allow for a uniform experience by the participants of the study. User studies with professional streamers and real spectators could shed further insights on \ourtooln, but this could be difficult to control variables.

\section{Summary}

In this paper, we identify key challenges related to live-streaming a VR experience to many spectators. We introduce an additional role, the co-host, to facilitate the communication between the VR user and spectators. We propose the \ourtool system to provide an interface for the co-host, which has features like private text-audio controls, camera feed selections, and asymmetric interactive tools. Our user study finds that the co-host adds value to the live-streaming experience for both the VR user and the spectators by processing information from a scalable number of spectators. We hope our work can inspire further studies and systems to enable scalable user interaction in VR spectating systems.

\bibliographystyle{ACM-Reference-Format}
\bibliography{citation}

\end{document}